# DEVELOPMENT OF MULTIVARIATE ATTENTION LSTM MODEL FOR DYNAMIC LINE RATING FORECASTING

**A.S. Bandara[1*], S.J. Siriwardena [1], A Wijethunge [2], J Ekanayake [1]**
*[1]Department of Electrical and Electronic Engineering, University of Peradeniya, Peradeniya 20400, Sri Lanka*
*[2]Department of Material and Mechanical Technology, University of Sri Jayewardenepura, Homagama 10200 Sri Lanka*
**Correspondence E-mail: *anushkab@ee.pdn.ac.lk*

**Abstract:** As global fossil fuel reserves diminish, there's a growing impetus for nations to transition towards renewable energy sources. Sri Lanka, for instance, aims to generate 70% of its electricity from renewable sources by 2030. Achieving this target requires optimal use of the existing power transmission infrastructure, as expanding the grid is both time-consuming and expensive. Traditionally, Static Line Ratings (SLRs) are used to define line capacity, often resulting in underutilization. Dynamic Line Rating (DLR), which estimates line capacity in real time based on weather conditions, offers a more efficient solution. However, DLR prediction is highly sensitive to environmental variability and forecasting complexity. This study proposes a novel multivariate Long Short-Term Memory (LSTM) model enhanced with an attention mechanism for improved DLR forecasting. Unlike traditional models that treat weather variables independently, the proposed approach captures nonlinear interdependencies among key environmental features such as ambient temperature, cable temperature, wind speed, humidity, and solar irradiance. The attention mechanism dynamically prioritizes the most relevant inputs during forecasting, leading to improved performance. Experimental evaluation on real-world DLR data demonstrates that the proposed model achieves a prediction accuracy of 95.84%, surpassing the conventional LSTM model's 94.62%. This improvement highlights the model's superior ability to deliver accurate and robust DLR forecasts. The findings confirm that incorporating multivariate features with attention enhances forecasting precision, supporting more efficient transmission line utilization and higher renewable energy integration.

## 1. Introduction

As global fossil fuel reserves continue to diminish, there is a growing impetus for nations to transition toward renewable energy sources. In line with this global shift, Sri Lanka has outlined a national target to generate 70% of its electricity from renewable sources by 2030 under the Renewable Energy Resource Development Plan 2021–2026. Achieving this target, however, poses significant challenges, especially regarding the expansion and modernization of the power transmission and distribution infrastructure. Given the high capital and time investment associated with upgrading the existing grid, optimizing current infrastructure becomes a cost-effective and feasible solution.

A key opportunity in this domain lies in how transmission lines are rated and utilized. Conventionally, power systems employ Static Line Ratings (SLRs) to define the maximum current a transmission line can carry under worst-case assumptions typically high ambient temperature and low wind conditions. While this conservative approach ensures safety, it often results in underutilization of line capacity.

To address this, Dynamic Line Rating (DLR) has been introduced. DLR provides a real-time estimate of a transmission line's capacity based on prevailing weather conditions such as ambient temperature, wind speed and direction, and solar radiation (Fernandez et al., 2016; A. Douglass et al., 2019). Numerous studies have demonstrated that DLR can significantly enhance line utilization without compromising safety, thereby facilitating greater integration of renewable energy sources such as solar and wind power (Wallnerstrom, Huang and Soder, 2015).

However, the deployment of DLR in operational environments remains limited due to its high dependency on accurate weather data and forecasting. Most early studies utilized deterministic or single-variable forecast models. For example, Gao et al. proposed ForecastNet, a deep learning-based model for day-ahead DLR forecasting using a single environmental variable (Gao et al., 2023). Similarly, Dupin et al. employed quantile regression forests to incorporate uncertainty but evaluated each variable independently (Dupin, Michiorri and Kariniotakis, 2019). Madadi et al. also modeled stochastic weather behavior using Ornstein–Uhlenbeck processes, but lacked multivariate feature interaction (Madadi, Mohammadi-Ivatloo and Tohidi, 2020).

To overcome these limitations, recent research has explored more advanced multivariate deep learning approaches. Sun and Jin introduced a spatiotemporal weather-based probabilistic forecasting model that leverages data from multiple nearby weather stations (Sun and Jin, 2022). Kim et al. proposed D-LGCLSTM, a hybrid model combining long short-term memory (LSTM) networks with graph convolution to capture both temporal and spatial dependencies in the network (Kim, Dvorkin and Kim, 2024).
Building further on deep learning advances, Vatsal et al. developed a Bi-directional LSTM (Bi-LSTM) model to predict DLR 24 hours ahead using past weather data. The model significantly outperformed traditional LSTM in terms of mean squared error and robustness, and demonstrated that weather parameters like wind speed, solar irradiance, and ambient temperature exhibit nonlinear but strong correlations with ampacity (Vatsal et al., 2024). Their study confirmed that Bi-LSTM effectively captures sequential dependencies in both temporal directions, leading to superior short-term prediction results.
In a parallel development, Liu et al. proposed an EMD-BiLSTM-BO model to forecast micrometeorological parameters for DLR application on a 500 kV line in China (Liu et al., 2021). Their approach combines Empirical Mode Decomposition (EMD) to handle non-stationary data, Bayesian Optimization (BO) for hyperparameter tuning, and recursive multi-step prediction to support day-ahead forecasting. To mitigate risk from accumulated prediction errors, they incorporated a Gaussian conservative correction model and simulated thermal aging of conductors using an Arrhenius-Weibull model. This comprehensive framework highlighted the importance of combining prediction accuracy with system safety and asset longevity.
Motivated by the need to address the inherent nonlinearity and interdependence of environmental factors, this study aims to develop a multivariate Attention-LSTM based DLR forecasting model. Unlike existing models that treat input features independently, our approach specifically focuses on capturing the nonlinear interdependence of environmental parameters such as ambient temperature, cable temperature, wind speed, humidity, and solar irradiance. The proposed model contributes to improving the operational efficiency of transmission assets and supports the realization of renewable energy targets by unlocking latent hosting capacity in existing networks.

## 2. Methodology

### 2.1 Evaluation of Dynamic Line Rating

Dynamic Line Rating (DLR), in contrast, dynamically estimates the ampacity of a conductor in real time based on actual environmental and operational conditions, such as conductor temperature, wind speed, ambient temperature, solar radiation, and current loading. The ampacity of a conductor under DLR is computed by solving the heat balance equation (Wijethunga, Wijayakulasooriya and Ekanayake, 2015).

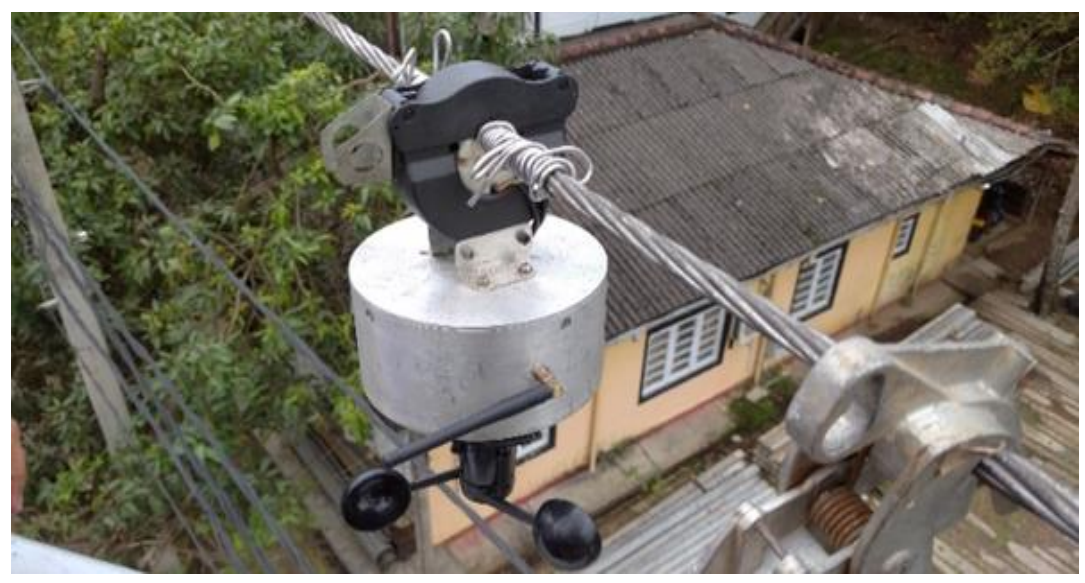

Figure 1: Wireless sensor node for DLR monitoring.

To enable accurate and low-cost DLR implementation on distribution lines, a wireless sensor node was designed based on the conductor-mounted measurement approach proposed in (Wijethunga et al., 2015). Figure 1shows a designed sensor node capable of measuring phasor current, conductor temperature, wind speed, humidity, and ambient temperature. These measurements are transmitted to a central server via a GPRS transmitter, where the DLR is calculated using an analytical model.
This study extends the functionality of the DLR system by introducing a data-driven forecasting framework. The objective is to predict short-term future DLR values using real-time and historical data captured by a sensor node to enhance the operational flexibility of the distribution network by providing predictive visibility into thermal line limits.

**2.2 Forecasting Framework**

The proposed scheme for predicting DLR precisely involves two main cases: a prediction utilizing only existing DLR data, and an enhanced prediction incorporating DLR data alongside meteorological variables. These cases are summarized in Figure 2 and detailed in the following subsections.

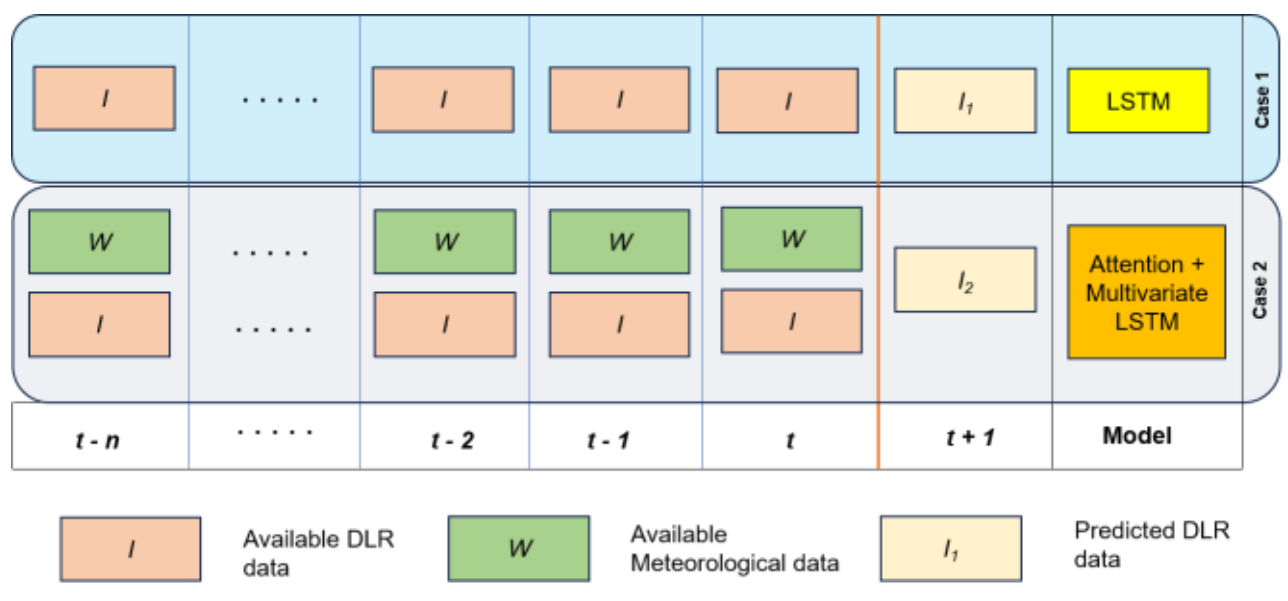

Figure 2: Summary of the proposed work.

### 2.2.1 Case 1

This initial step, depicted as Case 1 in Figure 2, demonstrates the process of DLR forecasting solely based on its past observations. A Long Short-Term Memory (LSTM) network is employed for this purpose. LSTM models are well-suited for sequential learning tasks due to their inherent ability to capture long-term dependencies within time series data. In this phase, for each time step t, the input to the LSTM is a scalar value I, representing the DLR at that specific time. For a sequence of n past time steps, the input would be a vector of DLR values $[I_{t-n}, I_{t-n+1}, \ldots\ldots, I_t]$. These historical DLR data available for a preceding time period are utilized as the training set for the LSTM model. The output of this case 1 is a DLR forecast ($I_1$), representing a single predicted DLR value for a future time point, serving as a baseline prediction without external influencing factors.

### 2.2.2 Case 2

Building upon the first step, Case 2 aims to improve DLR prediction by integrating additional relevant meteorological parameters. As illustrated in Case 2 of Figure 2, this phase utilizes a multivariate LSTM model enhanced with an attention mechanism. At each time step t, the input to this model is a vector comprising both available DLR data and critical meteorological variables. Specifically, the input vector $X_t$ for a given time step t is formulated as $[I_t, W_{t,temp}, W_{t,wind}, W_{t,humid}, W_{t,cable_temp}, W_{t,irrad}]$, where $I_t$ is the DLR value, and $W_{t,temp}$, $W_{t,wind}$ , $W_{t,humid}$ , $W_{t,cable_temp}$ and $W_{t,irrad}$ represent ambient temperature, wind speed, humidity, cable temperature and irradiance respectively at time t. For a sequence of n past time steps, the model processes a sequence of such vectors, $[X_{t-n}, X_{t-n+1}, \ldots.., X_t]$

The multivariate LSTM processes this combined dataset, recognizing the interdependencies between DLR and these meteorological factors. The attention mechanism, defined in Eq. (1), is incorporated to dynamically weight the prediction process. This allows the model to selectively focus on the most relevant information from the diverse input streams at each specific prediction instance. By leveraging this enhanced architecture, the model aims to produce a more precise DLR forecast ($I_2$), according to the complex influences of environmental conditions.

$$Attention\ (Q, K, V)\ = softmax\left(\frac{QK^T}{\sqrt{d_k}}\right)V \qquad (1)$$

### 2.3 Model Training and Evaluation.

The forecasting models are trained on 80% of the dataset and 20% for testing. The following performance metrics are computed to evaluate model accuracy.

#### 2.3.1 Mean Squared Error (MSE)

MSE is used to measure the root mean square difference between the predicted value and the actual (observed) value in a prediction problem.

$$MSE = \frac{1}{n}\sum_{i=1}^{n}(y1 - y2)^2 \qquad (2)$$

where n is the number of data points in the dataset and (y1 - y2) represents the difference between predicted and actual value. A lower MSE indicates that the model's predictions are closer to the actual values.

#### 2.3.2 Mean Absolute Error (MAE)

Mean absolute error is used to measure the average absolute difference between predicted values and actual values in a regression problem.

$$MAE = \frac{1}{n}\sum_{i=1}^{n}|y1 - y2| \qquad (3)$$

where n is the number of data points in the dataset and |y1 - y2| represents the absolute difference between predicted and actual value. MAE provides a more straightforward measure of prediction accuracy compared to MSE because it doesn't square the errors.

#### 2.3.3 R-squared ($R^2$)

R squared is also known as the coefficient of determination. It measures the proportion of variance in the target variable (dependent variable) that can be predicted from the independent variables attribute used in the regression model. An $R^2$ value close to 1 indicates that the model explains a larger proportion of the variance in the target variable, indicating a better fit. An $R^2$ value close to 0 indicates that the model does not explain much variance and may not be suitable for fitting.

## 3. Results and Discussion

In this study, a 33kV overhead distribution line located in Theldeniya, Sri Lanka, was considered. This line employs ACSR conductors, commonly used in networks across the country. These conductors typically have a cross-sectional area of approximately 35mm2 and support static line ratings (SLR) in the range of 95A to 125A, depending on ambient conditions and span configuration.

The dataset used in this analysis was collected using a DLR sensor node shown in Figure 1, mounted on the selected distribution line. The corresponding trends and relationship between environmental parameters and line ratings of sample observations are illustrated in Figure 3. The analysis indicated a strong correlation between solar irradiance and other parameters such as cable temperature, ambient temperature, and humidity, which in turn influence the conductor's current-carrying capacity. However, due to the non-linear nature of thermal exchange mechanisms, including convective and radiative cooling, the effect of each environmental factor on the line rating varies significantly across time. Therefore, relying on a single parameter, such as ambient temperature, is insufficient for accurate DLR estimation. It is essential to consider all relevant environmental conditions in forecasting models to ensure precise and reliable prediction of line ratings.

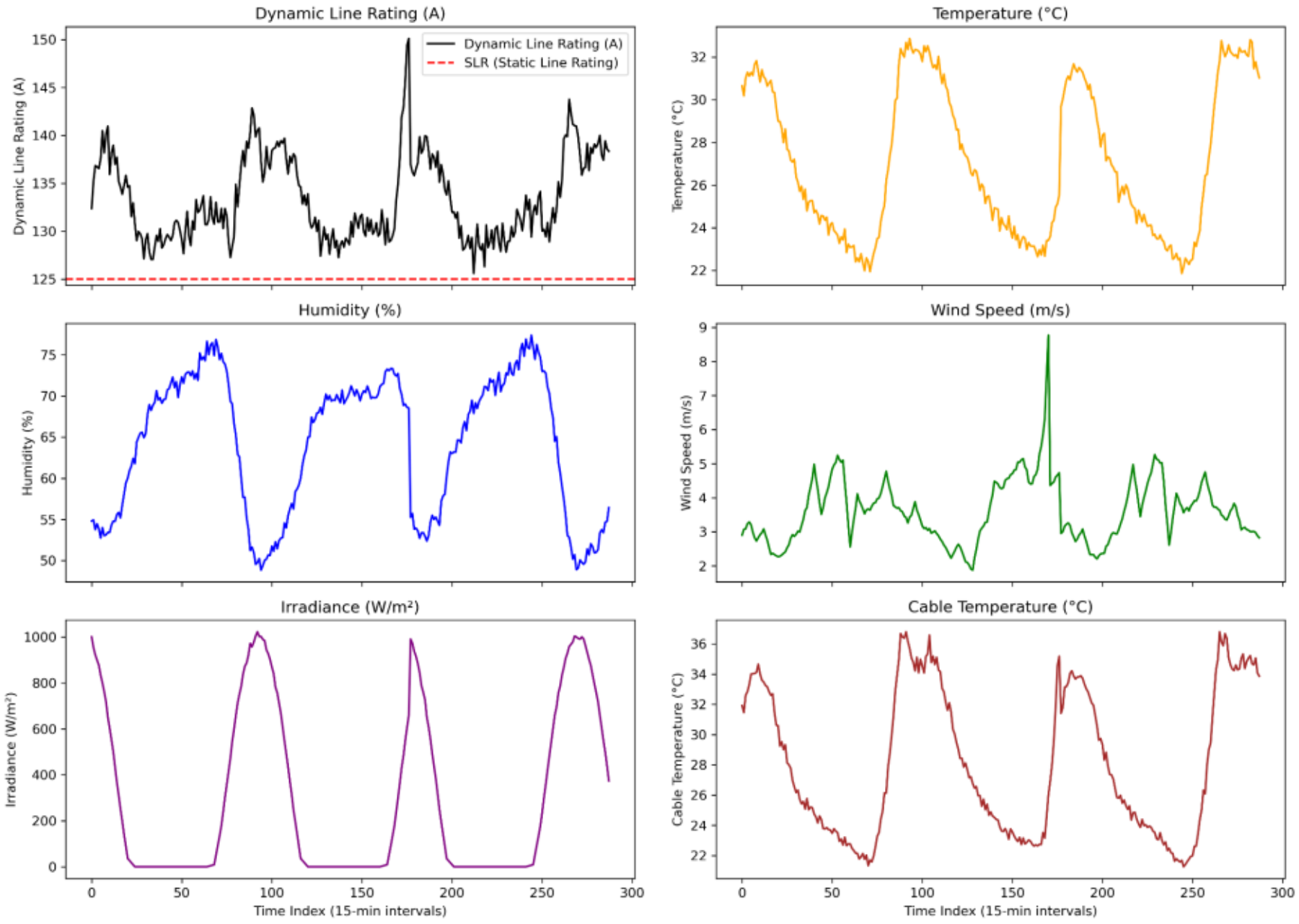

Figure 3: Distribution of DLR, temperature, humidity, wind speed, irradiance, and cable temperature

Table 1: MSE, RMSE and R-squared value of two models

| **Model** | **MSE** | **MAE** | **$R^2$** |
|---|---|---|---|
| Case 1 | 9.4974 | 2.4570 | 0.9462 |
| Case 2 | 7.3437 | 2.173 | 0.9584 |

The research leverages historical weather data with a high temporal resolution (15 minutes) to forecast DLR. The dataset was strategically split into training and testing sets to ensure robust model training on a representative sample and objective evaluation on unseen data. The performance of the two LSTM models, Case 1 and Case 2, is quantitatively assessed in Table 1, presenting their MSE, MAE and R-squared values for 15 minutes ahead prediction.

As evidenced by Table 1, the Multivariate LSTM model (Case 2) consistently outperforms the LSTM model (Case 1) across all metrics. Specifically, the Multivariate model achieved a lower MSE (7.3437 vs. 9.4974) and MAE (2.1730 vs. 2.4570), indicating a reduced prediction error. Furthermore, its higher R2 value (0.9584 vs. 0.9462) demonstrates that the Multivariate model explains a greater proportion of the variance in the DLR data, signifying a better fit.

Visually, the superior performance of the Multivariate LSTM model is clearly depicted in Figure 4.the plot distinctly illustrates how the predicted data of Multivariate LSTM model (Case 2) line align more closely with the testing data compared to the LSTM model (Case 1).

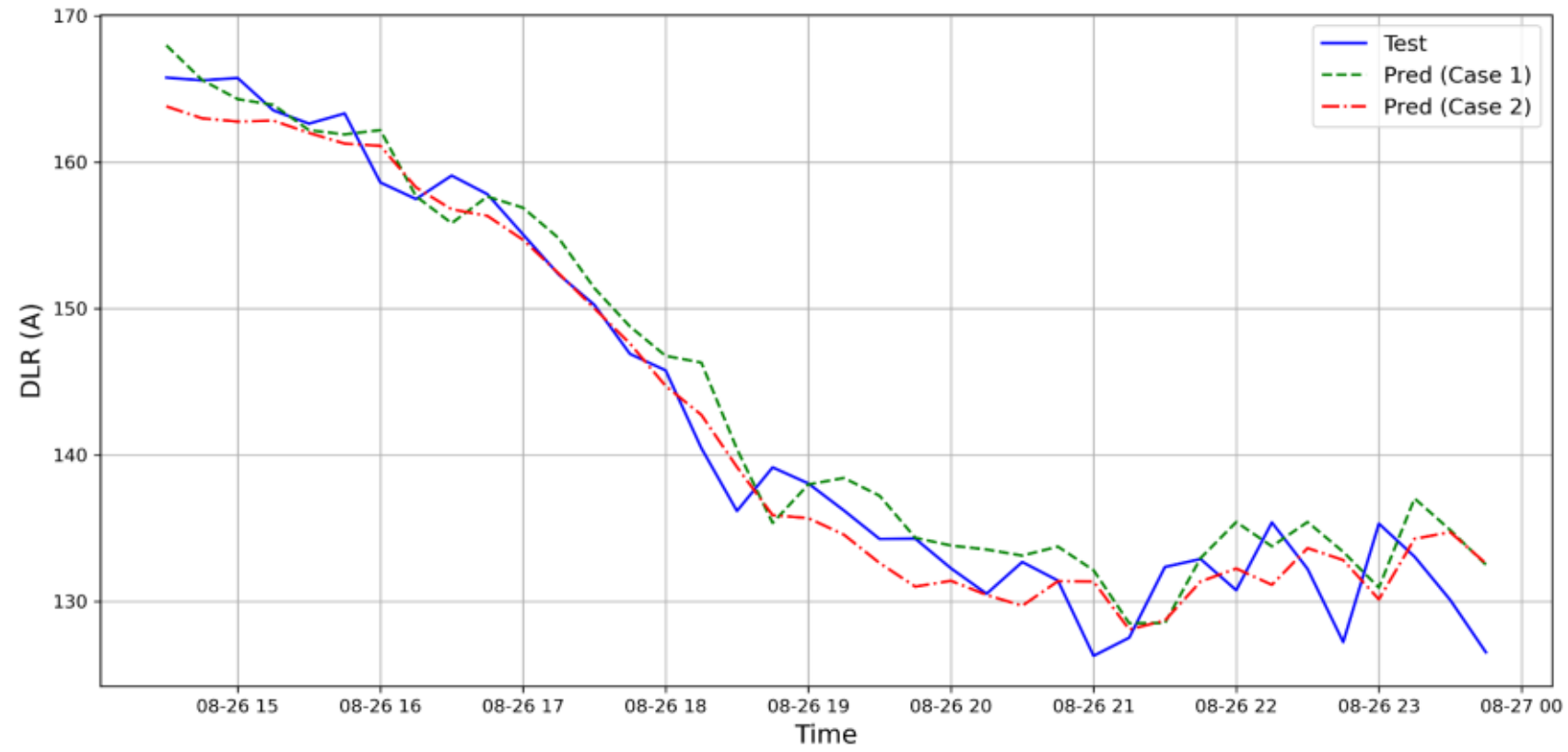

Figure 4: Actual vs Predicted value comparison between Case 1 and Case 2 LSTM models

## 4. Conclusion

This paper introduced an attention-based multivariate LSTM model to forecast Dynamic Line Rating (DLR) more accurately by learning complex relationships between environmental parameters. The model's ability to selectively focus on the most influential inputs enabled better handling of nonlinear interactions among variables such as temperature, wind speed, humidity, cable temperature and irradiance. Comparative results demonstrated that the proposed model outperformed a conventional LSTM baseline, achieving a higher prediction accuracy of 95.84% versus 94.62%. This gain reflects not only improved forecasting performance but also the model's practical potential for real-time grid operation and planning. By enabling more precise estimation of transmission line capacity, the proposed approach can significantly enhance grid flexibility and support higher levels of renewable energy integration—key factors in modernizing power systems for sustainable energy futures.

## Acknowledgements

The authors wish to extend their profound gratitude to the Department of Electrical and Electronic Engineering, Faculty of Engineering, University of Peradeniya, Sri Lanka, for facilitating this research.